\documentstyle[aps,graphics,epsfig,prd,manuscript]{revtex}
\begin{document}
\rightline{\small{\today}}
\vspace{5mm}
\begin{center}

{\bf The Statistical Analysis of Gaussian and Poisson Signals Near
Physical Boundaries}

\vspace{5mm}

Mark Mandelkern and Jonas Schultz\\
Department of Physics and Astronomy\\
University of California, Irvine, California 92697
\vspace{5mm}
\end{center}

\begin{abstract}

We propose a construction of frequentist confidence intervals that is
effective near unphysical regions and unifies the treatment of two-sided
and upper limit intervals.  It is rigorous, has coverage, is
computationally simple and avoids the pathologies that affect the
Likelihood Ratio and related constructions. Away from non-physical
regions, the results are exactly the usual central two-sided intervals.
The construction is based on including the physical constraint in the
derivation of the estimator, leading to an estimator with values that are
confined to the physical domain.

\vspace{5mm} 
\noindent 
PACS number(s):
02.50.Cw, 02.50.Kd, 02.50.Ph, 06.20.Dk 
\end{abstract} 

\newpage 
\section{Introduction}
\hspace{5mm}

\begin{figure}
\scalebox{0.8}[0.8]{\includegraphics[1cm,3cm][22cm,22cm]{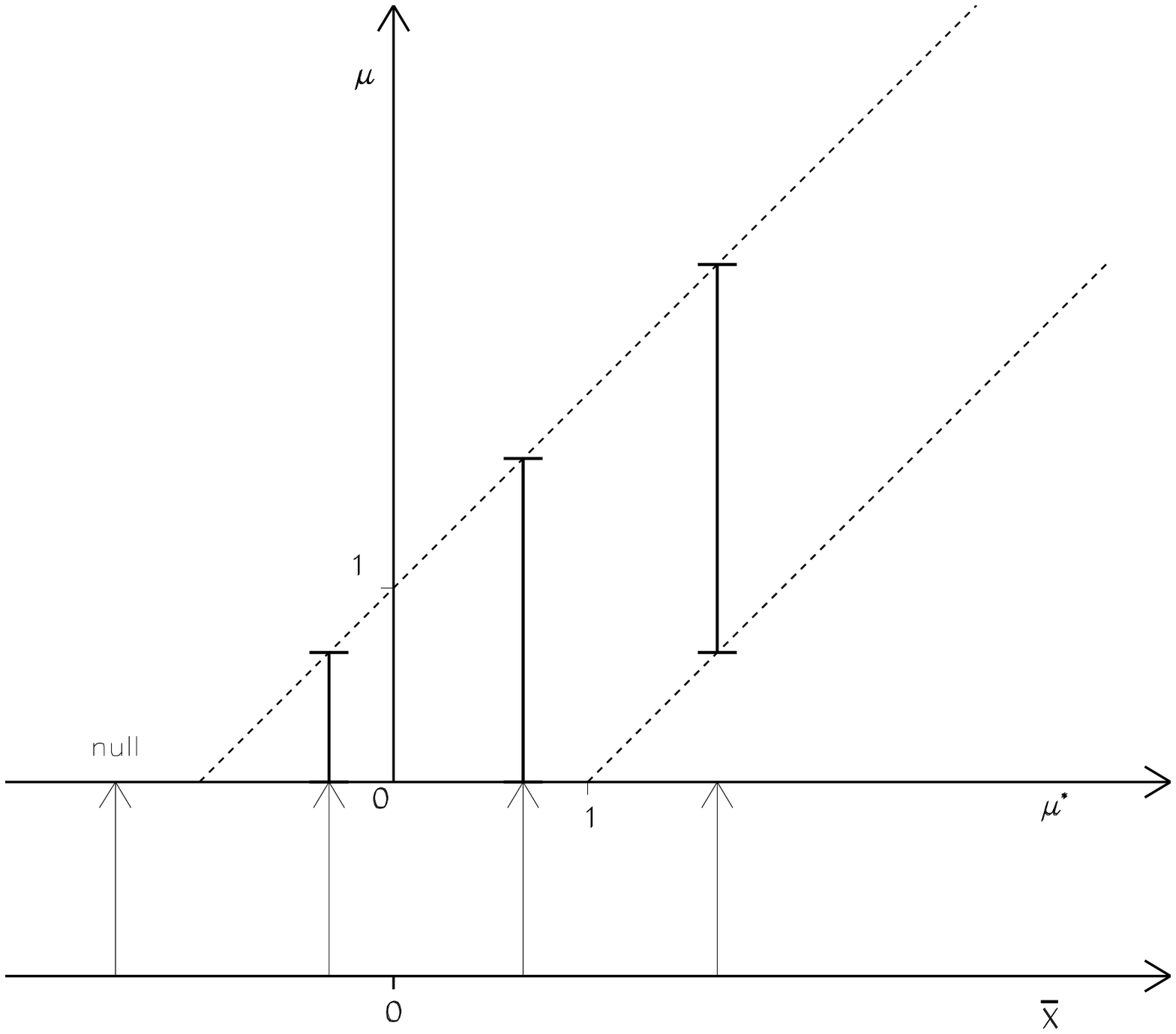}}
\caption{Confidence belt, in the usual construction, giving 68.27\%
central confidence intervals for the unknown mean of a Gaussian with
variance $\sigma^2$, in units of $\sigma_N= \sigma/\sqrt(N)$, where ${\bar
x}$ is the sample mean of $N$ measurements.}
\label{fig:classical_normal}
\end{figure}

\begin{figure}
\scalebox{0.90}[0.90]{\includegraphics[1cm,0.5cm][19cm,19cm]
{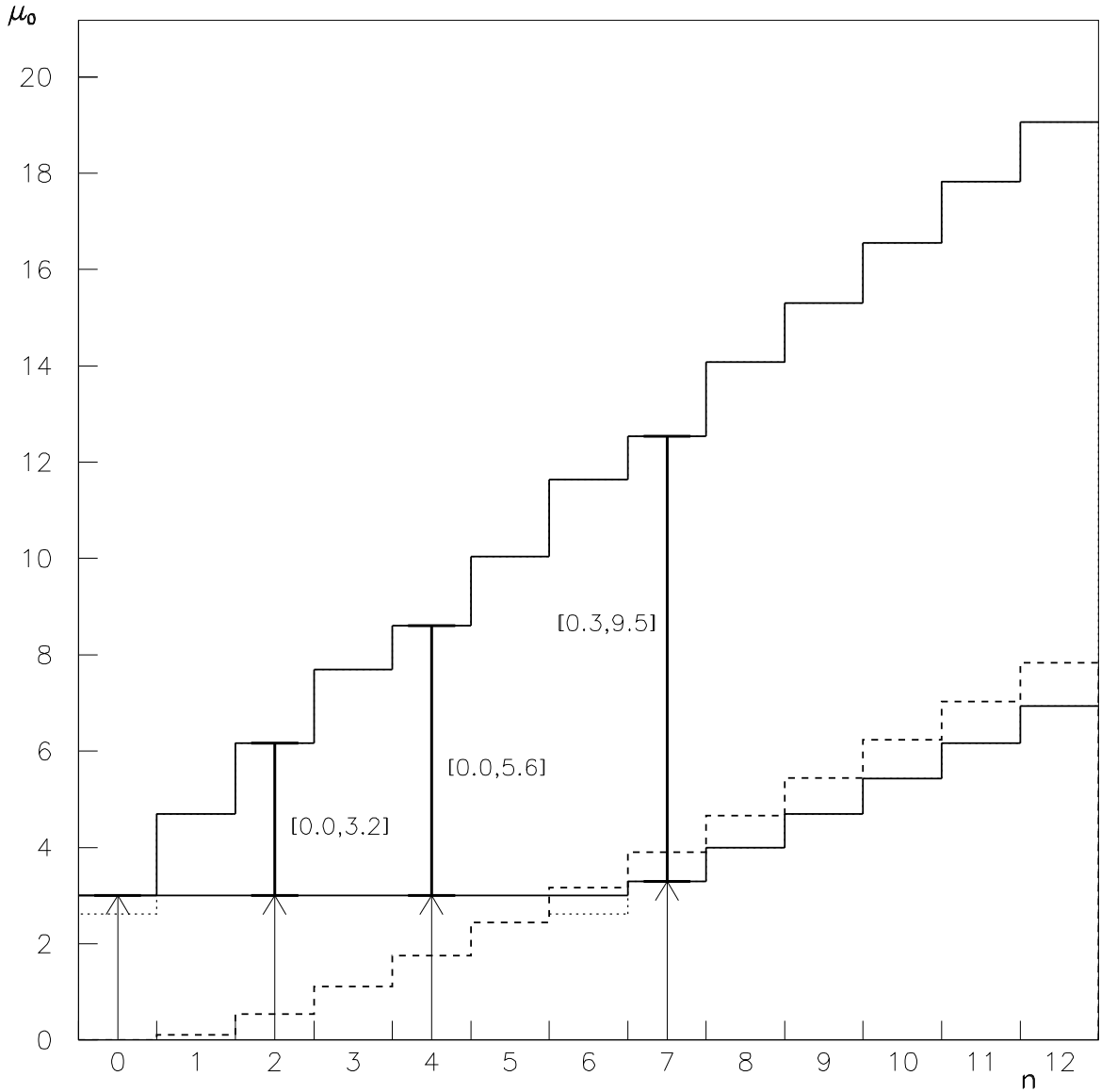}}
\caption{The classical construction of the 90\% central confidence belt
(solid) for unknown non-negative Poisson signal $\mu$ in the presence of a
Poisson background with known mean b taken to be 3.0, where $n$ is the
result of a single observation. Here $\mu_0=\mu+b$ is the parameter
representing the mean of signal plus background. For $n=0$ the confidence 
interval is null.} 
\label{fig:classical_poisson}
\end{figure}

Obtaining confidence intervals near physical boundaries is a long-standing
problem. Experiments designed to detect a non-zero neutrino mass by
observing neutrino oscillation or to detect a small resonance signal in
the presence of background are examples in which a negative result may be
obtained for a quantity that is intrinsically positive. The difficulty
arises when the estimate for the Gaussian or Poisson mean, as obtained
from the data, is near or beyond the physical boundary, in which case the
standard (classical) result of Neyman's construction is an unphysical or
null interval as illustrated in Fig.s \ref{fig:classical_normal} and
\ref{fig:classical_poisson}. 

For the Gaussian case, Fig. \ref{fig:classical_normal}, one obtains
central confidence intervals for the mean $\mu$ constrained to be
non-negative, using the sample mean ${\bar x}$ as the estimator for $\mu$. 
${\bar x}$ sufficiently negative leads to the null interval.  Despite the
fact that the construction has coverage $\alpha$, which means that, for
any given true mean, the confidence interval includes that value with
probability $\alpha$, the null interval {\it cannot} contain a true
non-zero mean. It is necessarily one of the measured intervals that, with
probability $1-\alpha$, fail to contain the true mean.  Even the non-null
intervals obtained by this method for some negative values of the
estimator are unphysically small in that, {\it for most possible (true)
means}, the confidence interval does not contain the true mean.

The other difficult case, illustrated in Fig. \ref{fig:classical_poisson},
is that of Poisson distributed data with unknown signal mean $\mu \geq 0$,
in the presence of a background with known mean $b$; $n$ is the result of
a single observation. For $n<b$ the interval for $\mu$ is unphysically
small. For sufficiently small $n$ the interval is null. The implausibility
of the resulting intervals is well illustrated by the example shown. For a
background-free ($b=0$) experiment that measures zero events($n=0$), the
90\% upper limit for $\mu$ is 2.62, for the explicit construction
exhibited in Fig. \ref{fig:classical_poisson}\cite{fn:1}. For an
experiment with known mean background $b=3.0$ that measures 0(1) events,
the upper limit for $\mu$ is 0(1.7). Thus the poorer experiment has the
potential to yield a much smaller (but not believable) upper limit. 

When the estimator takes on a value near or beyond the physical limit, we
have information greater than that available when no boundary is present
since we know {\it a priori} that the true value is not beyond the
boundary. For the Gaussian case, where the confidence intervals are of
fixed length for measurements away from the boundary, we expect smaller
confidence intervals for measurements near or beyond the boundary.  The
classical construction gives this feature. We also know that an estimate
for the parameter beyond the physical limit is relatively improbable. The
flaw in the standard classical method is that increasingly improbable
estimates lead to increasingly small and ultimately null confidence
intervals. One cannot accurately estimate a parameter by making an
extremely improbable observation. The best result for the determination of
a parameter should follow from the most probable measurement and,
arguably, the smallest confidence interval should be obtained for that
observation, i.e. ${\bar x}=\mu$ for the Gaussian case and $n=b+\mu$ for
the Poisson case.

\section{ Previously Suggested Methods for Obtaining Improved Confidence
Intervals}
\hspace{5mm}

A number of suggestions have been made for estimating believable
confidence intervals for bounded parameters. In the Review of Particle
Properties \cite{bib:PDG96,bib:PDG98}, the Particle Data Group suggests
several options for revising the intervals described above to make them
conservative, leading to overcoverage for small true values, and also
discusses the use of ``Bayesian upper limit(s), which must necessarily
contain subjective feelings about the possible values of the parameter".
 
Recently, several authors have suggested the use of different {\it
selection principles} for the construction of intervals. In the Neyman
construction, the confidence belt depends both on the properties of the
estimator and a selection principle.  The Neyman construction can be
simply described by means of a plot containing values of the estimator on
the abscissa and values of the parameter on the ordinate. According to
some prescription, i.e. the selection principle, one selects, for any
given value of the parameter, a horizontal interval corresponding to a
designated probability (the {\it coverage}) as determined by the sampling
distribution of the estimator. The region mapped out in this way for all
values of the parameter constitutes the {\it confidence belt}. After an
experiment is performed, yielding a specific value for the estimator, the
corresponding confidence interval for the parameter, with the designated
coverage, is the vertical interval contained in the confidence belt at
that value of the estimator. The most commonly used selection principles
(for coverage $\alpha$) are {\it central} (probability $\alpha$ within the
belt and equal probabilities on either side) and {\it one sided} (0 lower
limit and thus probability $\alpha$ to the left of ${\bar x_{upper}}$). 
One has the freedom to depart from the usual selection principles by, for
example, invoking a selection which makes the confidence belt as narrow as
possible \cite{bib:cg}. 

In recently suggested modifications, Ref. \cite{bib:fc} addresses both the
Gaussian and Poisson cases while Ref. \cite{bib:giunti} deals only with
the Poisson case. These approaches employ {\it ordering principles} for
the selection, i.e. rules which order the outcome probabilities before
aggregating to give total probability $\alpha$ for each value of the
parameter. In particular, the ordering is based on the Likelihood Ratio
Construction \cite{bib:cox} (and a variant), where the physical constraint
on the parameter space is used in the computation. These constructions
produce finite confidence intervals for all values of the classical
estimator and also achieve the admirable unifying feature that one need
not decide beforehand whether to set a confidence interval or a confidence
bound. However, the intervals obtained are small for {\it improbable}
values of the estimators and share with the classical central construction
the difficulty that, for a quite improbable value, the confidence interval
approaches the null interval. Thus, for the Gaussian case, a very negative
measured value yields a very small confidence interval with lower limit
zero.  Table X of Ref.  \cite{bib:fc} gives the confidence interval for
the (non-negative) Gaussian mean $\mu$ for measured value $x_0$. For
measured value $-3.0$ (unit variance assumed), the 68.27\% confidence
interval is [0.00, 0.04].  Despite the fact that this construction has
68.27\% coverage, the confidence interval derived from this measurement
does not contain the true value {\it for most possible true values} of the
Gaussian mean (excepting those in [0.00, 0.04]) that can lead to the
measurement. The resulting confidence interval is unphysically small. It
does not imply, in the words of the authors, a high ``degree of belief"
that the true value is within the interval. Our construction, which is
described below, yields [0, 1.0]. 

For the Poisson example cited above, of an experiment with known mean
background $b$ of 3.0 and a single observation yielding $n=0$, the 90\%
interval for the signal $\mu$ given by Ref.s \cite{bib:fc} and
\cite{bib:giunti} are [0, 1.08] and [0, 1.86] respectively, smaller than
the interval given for $n=0$, $b=0$ of [0, 2.44]. Ref. \cite{bib:giunti}
emphasizes that the reason for obtaining small upper limits for $n<b$ is
not increased sensitivity to the signal but just that fewer background
events than expected are observed, and views it as ``an undesirable
feature from the physical point of view" for the upper limit to decrease
as $b$ increases. Our construction, described below, yields [0, 2.62] for
the $b=0$ case and [0, 4.69] for the $b=3.0$ case, thus a larger rather
than smaller interval for 0 events measured when background is present. Of
the constructions discussed here, ours is the only one where the upper
limit increases rather than decreases as $b$ increases for fixed $n$.

In recognition of the problem of unphysical intervals, Ref. \cite{bib:fc}
introduces the concept of ``sensitivity" to handle cases in which the
measurement is less than the estimated background and the confidence
interval is suspect. This, however, requires quoting a second value, a
characteristic of the experiment itself, in addition to the interval
quoted. No substitute interval is offered.

The authors of Ref. \cite{bib:rw} construct confidence intervals for the
Poisson case. They point out that the observation $n=0$ implies that zero
signal is seen, thus the estimate for $\mu$ (zero) is independent of $b$.
They argue, therefore, that the confidence interval for $\mu$ for n=0 must
be independent of $b$. Extending the argument, they note that for any
observation $n$, one has observed a signal $n$ from the Poisson pdf
$p(n;\mu+b)$ {\it and} at most a background $n$. Thus they formulate a
method of obtaining confidence intervals based on the conditional
probability to observe $n$ given a background $\leq n$ and obtain the
desired result for $n=0$ and approximately the classical confidence
intervals for $n>b$. While they identify their method as an ordering
principle, it is not one in the same sense as Ref.s \cite{bib:fc} and
\cite{bib:giunti}. The confidence belt is not constructed from the
sampling distribution of an estimator and hence does not have coverage in
the usual sense.  The method gives intervals that are intuitively more
satisfying as measures of confidence. However, because the method does not
provide coverage, one cannot precisely state the probability that the
interval encloses the true value.

Although the intervals determined by the method of \cite{bib:rw} do not
have coverage, they can be easily modified so that they do, by
restructuring the confidence belt, retaining the lower limit and adjusting
the upper limit so that all horizontal intervals contain probability
$\alpha$. If one thus modifies the construction, the procedure represents
another selection principle applied to the Poisson pdf for the sample
mean.  For $n=0$ independent of $b$, this method gives a 90\% upper limit
of 2.42.

\section {Frequentist vs Bayesian Confidence Intervals}

The methods of Refs. \cite{bib:fc} and \cite{bib:giunti} are frequentist,
as they are constructed from the sampling distribution of an estimator, in
this case the sample mean, and have coverage by construction. However any
estimator may be chosen for the Neyman construction. The method used to
choose the estimator is arbitrary. The estimator may be a guess, or
arrived at by the usual techniques of moments or the Maximum Likelihood
Method. Although it is in general desirable for an estimator to be
sufficient and unbiased \cite{bib:eadie}, it need not have these
properties, so long as it possesses other desirable features, e.g.  gives
an appropriate point estimate of the parameter of interest and leads to
confidence intervals that are restrictive and believable from a physical
point of view. Coverage is guaranteed by construction. 

Bayesian confidence intervals are constructed from the Bayesian posterior
density, which is interpreted as the probability density for the unknown
parameter. A selection principle is again needed to specify the parameter
interval containing the designated probability. The Bayesian procedure for
confidence intervals does not guarantee coverage because it is not
obtained from the probability density of a statistic or random variable
and can be criticized for the subjectiveness inherent in establishing the
required Bayesian prior. For a discussion of Bayesian methods, the reader
is referred to Ref.  \cite{bib:lehmann}. Our interest is in a frequentist
method, as described in the following section. 

\section{Intervals Based on an Estimator Derived From A Likelihood
Function that Contains the Physical Constraints}

\hspace{5mm}

The authors cited above have focused on modifying the selection principle
to make the confidence intervals more believable. However the reason that
their constructions lead to unphysically small confidence intervals near
the boundary of a physical region is that the method used to obtain the
estimator does not take into account the physical constraint on the
parameter of interest and the resulting estimator is thus the same as if
there were no boundary. Even though that estimator is efficient, it is
appropriate for a problem other than the one under consideration. 

We propose a frequentist method and use the Maximum Likelihood Method to
derive the estimator employed. Among methods for determining estimators,
the Maximum Likelihood Method is preferred in that if a consistent
estimator exists, the method will produce it
\cite{bib:arley,bib:cramer,bib:eadie}. The Likelihood Function chosen
explicitly contains the physical constraint and leads to an estimator with
values within the physical domain that is appropriate for the problem. The
confidence intervals obtained consequently from the sampling distribution
of the estimator have coverage by construction, are more physical and
support a higher degree of belief that the parameter of interest lies
within the interval. 

This method follows classical estimation theory; the only new element is
that the Likelihood Function explicitly excludes non-physical true values. 
The determination of the estimator, its sampling distribution and the
confidence intervals follow directly without further assumptions. We
emphasize that the procedure we are following is not Bayesian and that the
exclusion of non-physical true values is not equivalent to a uniform
Bayesian prior for the physical region any more than the usual
unconstrained Likelihood Function is viewed as containing a uniform
Bayesian prior for the entire domain. 

\subsection{Gaussian variates}
\hspace{5mm}

We assume that $x$ is normally distributed with non-negative mean $\mu$ and
variance $\sigma^2$.

\begin{equation}
f(x|\mu)=\frac{1}{\sqrt{2\pi}\sigma}
exp\left[-\frac{(x-\mu)^2}{2\sigma^2}\right]
\end{equation}

The likelihood function, when there are $N$ measurements $x_1, x_2,
....x_n$, is:

\begin{equation}
L(\mu) = \prod^{N}_{i=1}f( x_i|\mu)\theta(\mu)
\end{equation}

\begin{equation}
w(\mu)=lnL(\mu)=\sum^{N}_{i=1}\left(-\frac{(x_i-\mu)^2}{2\sigma^2}\right)-
Nln(\sqrt{2\pi}\sigma)+ln\theta(\mu)
\end{equation}
where $\theta(\mu)$ is a step function; $\theta(\mu)=0$ for $\mu<0$,
$\theta(\mu)=1$ for $\mu\geq0$.  The estimator for $\mu$, which we denote
by $\mu^*$, is the function of the measurements, $\mu(x_i)$, that
maximizes $w$. Since $w=-\infty$ for $\mu<0$, $\mu^*$must be $\geq$ 0. We
set
\begin{equation}
\frac{dw}{d\mu}=\sum^{N}_{i=1}\frac{x_i-\mu}{\sigma^2} =0
\end{equation}
For the sample mean $ {\bar x}\equiv\frac{1}{N}\sum^{N}_{i=1}x_i \geq 0$,
$\mu^*={\bar x}$. For ${\bar x}<0$, $\frac{dw}{d\mu}<0$ for all $\mu\geq
0$, so the maximum of $w$ is at $\mu^*=0$.
${\bar x}$ has a normal distribution with mean $\mu$ and variance
$\sigma_N^2=\sigma^2/N$. The probability density function for $\mu^*$ is
normal with the usual normalization for $\mu^*>0$ and a delta function at
$\mu^*=0$ normalized to the remaining probability

\begin{equation}
P_0(\mu) \equiv \frac{1}{\sqrt{2\pi}\sigma_N}
\int_{-\infty}^0 exp\left[-\frac{(x-\mu)^2}{2\sigma_N^2}\right]dx
=\frac{1-erf(\mu/\sqrt{2}\sigma_N)}{2}
\end{equation}
Thus the probability density function for $\mu^*$ is given by:

\begin{equation}
P(\mu^*|\mu)= \frac{1}{\sqrt{2\pi}\sigma_N}
exp\left[-\frac{(\mu^*-\mu)^2}{2\sigma_N^2}\right]
+\delta(\mu^*)P_0
\end{equation}
The mean and variance of $\mu^*$ are given by:

\begin{equation}
E(\mu^*)=
\frac{\sigma_N}{\sqrt{2\pi}}
exp\left[-\frac{\mu^2}{2\sigma_N^2}\right]
+\mu(1-P_0)
\end{equation}

\begin{equation}
V(\mu^*)=(\sigma_N^2+\mu^2P_0)(1-P_0)-
\frac{\mu \sigma_N (1-2P_0)}{\sqrt{2\pi}}
exp\left[-\frac{\mu^2}{2\sigma_N^2}\right]-
\frac{\sigma_N^2}{2\pi}
exp\left[-\frac{\mu^2}{\sigma_N^2}\right]
\end{equation}
$E(\mu^*)$ approaches $\mu$ and $V(\mu^*)$ approaches $\sigma_N^2$ for $N$
large. For finite $N$, $E(\mu^*)$ does not equal $\mu$, so $\mu^*$ is a
consistent but not unbiased estimator for $\mu$. It is, however,
asymptotically unbiased. From Estimation Theory
\cite{bib:arley,bib:cramer,bib:eadie} we know that {\it If the Likelihood
Equation has a solution $\mu^*$ which is a consistent estimator of $\mu$,
then $\mu^*$ is asymptotically normally distributed with a mean of $\mu$
and a variance of $\left[-N E(d^2ln f(x|\mu)/d\mu^2)\right]^{-1}$}.
$V(\mu^*)$ equals 0.34$\sigma_N^2$ at $\mu$=0, monotonically increasing to
$\sigma_N^2$ at large $\mu$. For finite $N$, $V(\mu^*)$ is smaller than
$\sigma_N^2$.

Nevertheless, $V(\mu^*)$ satisfies the usual Cramer-Rao inequality
\cite{bib:eadie}
\begin{equation}
V(\mu^*) \geq \frac{\left(\frac{dE(\mu^*)}{d\mu} \right)^2}{I_X}
\end{equation}
where $I_X$ is the Fisher Information, the usual measure of the
information contained in the measurements. One finds
$\frac{dE(\mu^*)}{d\mu}=1-P_0$ and $I_X=\frac{1}{\sigma_N^2}$ and
by explicit calculation one can show that

\begin{equation}
V(\mu^*)  \geq  (1-P_0)^2 \sigma_N^2
\begin{array}{c} \\ \longrightarrow \\ ^{N\rightarrow \infty}
\end{array}
\sigma_N^2
\end{equation}

\begin{figure}
\scalebox{0.8}[0.8]{\includegraphics[1cm,3cm][22cm,22cm]{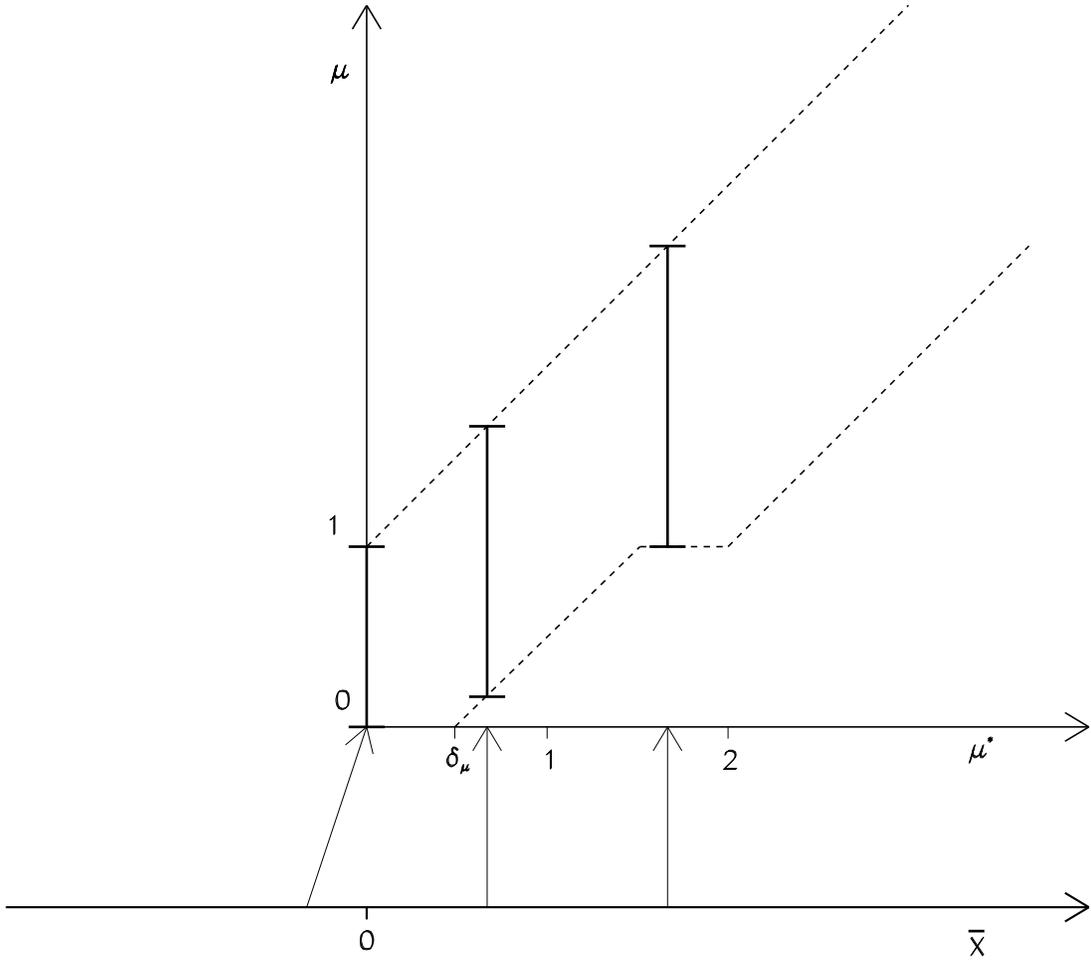}}
\caption{Confidence belt, in our construction, giving 68.27\% central
confidence intervals for the unknown mean of a Gaussian with variance
$\sigma^2$, in units of $\sigma_N = \sigma/\sqrt{N}$, where ${\bar x}$ is
the sample mean of $N$ measurements.  For ${\bar x}\leq 0$, $\mu^*=0$ and
the interval is [0, 1].  For $0 < {\bar x} \leq \delta_\mu$ the interval
is [0, ${\bar x}+ 1$]. For $\delta_\mu < {\bar x} \leq 1+ \delta_\mu$ it
is [${\bar x}-\delta_\mu, {\bar x}+1$]. For $1+\delta_\mu < {\bar x} \leq
2$ the interval is [$1, {\bar x}+1$] and for ${\bar x}>2$ we obtain the
usual central interval [${\bar x}-1$, ${\bar x}+1$]. $\delta_\mu =
0.475$.}
\label{fig:non-negative_normal}
\end{figure}

We note that $\mu^*$ does not satisfy the criteria for sufficiency.
However for the purpose of supplying a point or interval estimate for this
special case where there is a boundary, it contains all of the necessary
information. (For ${\bar x}<0$, the best estimate of $\mu$ is zero.)
We demonstrate the construction of the 68.27\% central confidence belt, in
units of $\sigma_N= \sigma/\sqrt{N}$, in Fig.
\ref{fig:non-negative_normal}. We invoke the Neyman construction and
select, for any given value of $\mu$, the ``central" interval of $\mu^*$
that contains 68.27\% of the $\mu^*$ sampling distribution. For $\mu$=0,
50\% of the $\mu^*$ probability distribution is associated with $\mu^*$=0.
The remaining 18.27\% of the 68.27\% belt is contained in the $\mu^*$
interval between 0 and what we call $\delta_{\mu}$. A straightforward
calculation gives $erf(\delta_{\mu}/\sqrt{2})=2 \times 0.1827$, or
$\delta_{\mu}$=0.475.

As $\mu$ increases from 0 to 1, the upper endpoint of the 68.27\% interval
rises linearly with unit slope. For $\mu>1$, the central 68.27\% interval
in $\mu^*$ is $\mu-1\leq\mu^*\leq\mu+1$. It is the requirement of exactly
68.27\% coverage, and the fact that the finite probability associated with
$\mu^*=0$ must be taken into account, that introduces a discontinuity in
the central interval at $\mu=1$.

Once the confidence belt is constructed, as in Fig.
\ref{fig:non-negative_normal}, it follows from the Neyman method that
confidence intervals of $\mu$ with corresponding coverage can be read off
as vertical intervals of the belt for any measured ${\bar x}$. We need
only keep in mind that all ${\bar x}<0$ correspond to $\mu^*=0$.

In our formulation, the necessary``lift up" \cite{bib:PDG96} of the
estimate from an unphysical to a physical value and/or the raising of an
upper bound to a non-null value comes naturally from the estimator derived
from the Likelihood Function. In other approaches,
\cite{bib:PDG96,bib:fc} the ``lift-up" is obtained somewhat
arbitrarily by ad hoc procedures or by specifying an ordering principle.
The latter methods do not solve the problem that, in the words of Ref.
\cite{bib:PDG96},``in some (rare) cases it is necessary to quote an
interval {\it known to be wrong.}"

\subsection{Poisson variates with background}
\hspace{5mm}

We consider $n$ to be a single Poisson distributed variate with
non-negative signal mean $\mu$ and known mean background $b$. Let
$p(n|m)=m^ne^{-m}/n!$ denote the Poisson probability for obtaining the
measurement $n$ when the mean is $m$.  Then

\begin{equation}
f(n|\mu)= p(n|\mu+b)
\end{equation}

\begin{equation}
L(\mu) = f(n|\mu)\theta(\mu)
\end{equation}

\begin{equation}
w(\mu)=lnL(\mu)=nln(\mu+b)-(\mu+b)-ln(n!)+ln\theta(\mu)
\end{equation}
$\mu^*$ is the function of $n$ that maximizes $L$ and is thus the estimator
for $\mu$.  For $n>b$, $\mu^*=n-b$. For $n\leq b$, $\mu^*=0$. Thus the
estimator for $\mu$ is non-negative. The probability of $\mu^*$ for a
given $\mu$ is $P(\mu^*|\mu,b)=p(\mu^*+b|\mu+b)$ for $\mu^* >0$ and a
value at $\mu^*=0$ given by $\sum_{n\leq b} p(n|\mu+b)$. Rather than work
with the estimator $\mu^*$, it is more convenient to define an integer
estimator, $n^*$, such that $n^*=0$ for $n \leq b$ and $n^*=n-b^-$ for
$n\geq b$, where $b^-$ is the largest integer less than or equal to $b$.
Thus $n^*=\mu^*+(b-b^-)$.

\begin{figure}
\scalebox{0.90}[0.90]{\includegraphics[1cm,0.5cm][19cm,19cm]
{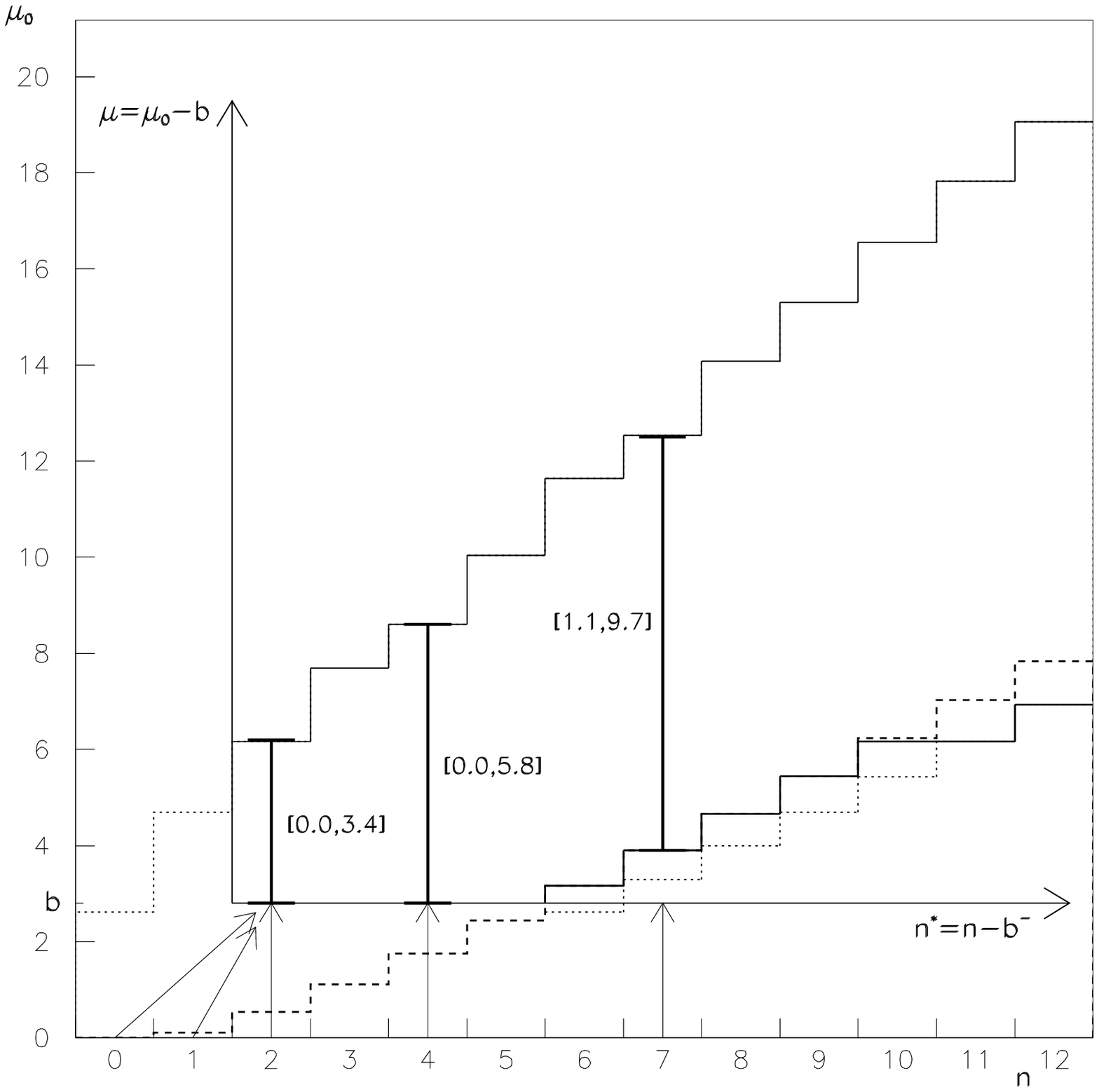}}
\caption{The 90\% central confidence belt (solid) for unknown non-negative
Poisson signal $\mu$ in the presence of a Poisson background with known
mean b taken to be 2.8, where $n$ is the result of a single observation. 
We show the confidence belt consisting of central intervals [$n_1(\mu_0)$,
$n_2(\mu_0)$] containing at least 90\% of the probability for unknown
Poisson mean $\mu_0$ in the absence of background (dotted) and the 90\%
one-sided belt consisting of intervals [0, $n_{os}(\mu_0)$](dashed). For
$\mu_0<2.62$, only one-sided intervals can be constructed.  For
$b=2.8$, $b^-=2$, $\mu'_0=6.2$, $n_{os}(b)=5$, and $n_{os}(\mu'_0)=9$ (see
text for definitions).  For $n\leq b$, the confidence interval for $\mu$
is [$0, c_2(b^-)-b=3.4$] and the examples given are for $n\leq 2$. For
$b<n\leq 5$, the interval is [$0, c_2(n)-b$] where the example given is
for $n=4$ and the interval is [0, 5.8]. For $5<n\leq 9$, the interval is
[$c_{os}(n)-b, c_2(n)-b$] where the example given is for n=7 and the
interval is [1.1, 9.7]; for $n=10$, the interval is [$\mu'_0-b$,
$c_2(n)-b$]; and for $n>10$, the interval is [$c_1(n)-b, c_2(n)-b$]. Here
[$c_1(m), c_2(m)$] is the Poisson central 90\% confidence interval and
$c_{os}(m)$ is the one-sided Poisson 90\% lower limit, both for a single
observation giving m in the absence of any known background.}
\label{fig:non-negative_Poisson}
\end{figure}

We demonstrate the construction of the 90\% confidence belt by means of an
example, shown in Fig. \ref{fig:non-negative_Poisson}, where the known
mean background $b$ is equal to 2.8. $b$ is chosen non-integer to
illustrate this slightly more complicated case. We also show the
confidence belt consisting of central intervals [$n_1(\mu_0)$,
$n_2(\mu_0)$] \cite{fn:2} containing at least 90\% of
the probability for unknown Poisson mean $\mu_0$ in the absence of any
known background (dotted) and the 90\% one-sided belt consisting of
intervals [0, $n_{os}(\mu_0)$](dashed).  Our 90\% confidence belt is
defined only for $\mu\geq 0$ and $\mu^*\geq 0$. We define a coordinate
system ($n^*,\mu$) by placing the ordinate $ \mu=0$ at $\mu_0=b$ and
choosing the integer abscissa value $n^*=0$ to coincide with $n=b^-$.

Let $\mu'_0$ be the largest value of $\mu_0$ such that
[$n_1(\mu_0),n_2(\mu_0)$] contains $b^-$. (In the example given,
$\mu'_0=6.2$, corresponding to a value of $\mu=6.2-2.8=3.4$, and
$n_{os}(\mu'_0)=9$.) For $0\leq\mu\leq\mu'_0-b$ (i.e. 
$b\leq\mu_0\leq\mu'_0$), the 90\% horizontal interval is
[$b^-,n_{os}(\mu_0)$]. For $\mu>\mu'_0-b$ (i.e. $\mu_0>\mu'_0$), the 90\%
horizontal interval is [$n_1(\mu_0), n_2(\mu_0)$]. The resulting
confidence belt is shown in solid lines. The set of joined horizontal and
vertical line segments is simple and continuous and no compensatory
remedies are required. To obtain the 90\% confidence intervals for $\mu$,
given a measurement $n$, we need simply find the appropriate vertical
interval from the plot. By the Neyman construction, it has $\geq$ 90\%
coverage. 

Let [$c_1(m), c_2(m)$] denote the usual (i.e. in the absence of known
background) Poisson 90\% confidence interval for the mean, $\mu_0$, for m
observed events (the dotted horizontal lines in Fig.
\ref{fig:non-negative_Poisson}) . Also, let $c_{os}(m)$ denote the usual
90\% one-sided lower limit for m observed events (the dashed horizontal
lines in Fig. \ref{fig:non-negative_Poisson}). Then for $n\leq b$, $n^*=0$
and we obtain the upper limit for $\mu$ of $c_2(b^-)-b$. For $b<n\leq
n_{os}(b)$ we obtain the upper limit $c_2(n)-b$; for $n_{os}(b)<n\leq
n_{os}(\mu_0')$ we obtain the interval [$c_{os}(n)-b, c_2(n)-b$]; for
$n=n_{os}(\mu'_0)+1$ we obtain the interval [$\mu'_0-b$, $c_2(n)-b$] and
for $n> n_{os}(\mu_0')+1$ we obtain the interval [$c_1(n)-b, c_2(n)-b$].
We note that any Poisson interval with known background can be obtained
from a single figure or table.

It is straightforward to generalize to the case of $N$ independent
measurements. For measured mean $\bar{n}\geq b$, $\mu^*=\bar{n}-b$. For
$\bar{n}<b$, $\mu^*=0$. The probability for $\mu^*$ is Poisson for
$\mu^*>0$ plus a value at $\mu^*=0$ normalized to the remaining
probability.

\begin{equation}
P(\mu^*|\mu,b)_{\mu^*>0}=p(N(\mu^*+b)|N(\mu+b))
\end{equation}
\begin{equation}
P(0|\mu,b)=\sum_{m\leq Nb}p(m|N(\mu+b))
\end{equation}
In this case we can find the confidence interval for $\mu^*$ by relabeling
the axes in Fig. \ref{fig:non-negative_Poisson} as follows:
$n\rightarrow N\bar{n}$, $\mu_0\rightarrow N\mu_0$, $n^*\rightarrow Nn^*$,
$\mu\rightarrow N\mu$, and the origin of the inner coordinate system is
$(Nb^-, Nb)$.

\section{Mass squared of the electron neutrino}
\hspace{5mm}

As an example we obtain the 68.27\% confidence interval for the mass
squared of the electron neutrino, disregarding the possibility that the
source of negative measurements is physics (fitting to the wrong function)
rather than statistical variation. Using the measurement quoting the
smallest error, that of Ref. \cite{bib:belesev} giving $-22\pm
4.8$~eV$^2$, and assuming Gaussian probability we obtain the interval [0,
4.8]. The classical Neyman interval is null and the interval offered by
Ref. \cite{bib:fc} is [0, 0.02] \cite{fn:3}.

\section{Conclusion}
\hspace{5mm}

We have demonstrated a rigorous method for obtaining frequentist
confidence intervals that incorporates the physical constraints of the
problem into the Likelihood Function, thus yielding an estimator that is
suitable to the presence of physical boundaries. Using a central ordering
principle, we obtain either upper limits or central intervals with a
smooth transition. The intervals are physical in that they support a high
degree of belief that the true value is within the interval, avoiding the
pathologies of null or unphysically small intervals and the consequent
possibility of obtaining a better result (smaller confidence interval) for
a worse experiment. The construction is not equivalent to the Likelihood
Ratio Construction which does not give satisfactory intervals near
unphysical regions.

\section{Acknowledgments} We thank Giovanni Lasio for assisting with the
figures and Ephraim Fischbach for suggesting the neutrino mass squared
example.

\begin{thebibliography}{99}

\bibitem{fn:1}We note that, depending upon the particular choice of
construction, the 90\% upper limit obtained for the case $b=0$, $n=0$ can
vary over a small range; e.g. the limit is 2.30 for a one-sided upper
limit construction, 2.44 for the methods of Ref.s \cite{bib:cg} and
\cite{bib:fc} and 2.62 for the construction presented here. 

\bibitem{bib:PDG96} R.M. Barnett {\em et al.}, Review of Particle Physics,
Phys. Rev. D{\bf 54}, 1 (1996). 

\bibitem{bib:PDG98} C. Caso {\em et al.}, Review of Particle Physics, Eur. 
Phys. J. {\bf C3}, 1 (1998). 

\bibitem{bib:cg} E. L. Crow and R. S.  Gardner, Biometrika {\bf 46}, 441
(1959). 

\bibitem{bib:fc} G. J. Feldman and R. D. Cousins, Phys. Rev. D{\bf 57},
3873 (1998).

\bibitem{bib:giunti}
C. Giunti, Phys. Rev. D{\bf 59}, 053001 (1999),\\
C. Giunti, Phys. Rev. D{\bf 59}, 113009 (1999).

\bibitem{bib:cox}D. R. Cox and D. V. Hinkley, Theoretical Statistics,  London: 
   Chapman and Hall, 1974.

\bibitem{bib:rw}B. P. Roe and M. B. Woodroofe, Phys. Rev. D{\bf 60},
053009 (1999).

\bibitem{bib:lehmann}E. L. Lehmann, Testing Statistical Hypotheses, 2nd
Ed., New York:Wiley, 1986.
 
\bibitem{bib:arley} N. Arley and K. R. Buch, Introduction to the Theory of
Probability and Statistics, New York : Wiley, 1950.

\bibitem{bib:cramer} H. Cramer, Mathematical Methods of Statistics,
Princeton: Princeton University Press, 1946.

\bibitem{bib:eadie} W. T. Eadie, D. Drijard, F. E. James, M. Roos and B.
Sadoulet, Statistical Methods in Experimental Physics, Amsterdam:
North-Holland, 1971.

\bibitem{fn:2}There is some arbitrariness in the choice of a
central interval for a discrete variate. We choose the smallest interval
such that there is $\geq$ 90\% of the probability in the center and
$\leq$5\%, but as close as possible to 5\%, on the right. The alternative
of requiring $\leq$5\%, but as close as possible to 5\%, on the left gives
slightly less symmetrical intervals. For the latter choice the 90\%
Poisson upper limit for $n=0$ is $\mu_0=3.0$ compared to $\mu_0=2.62$ for
our choice. For $\mu_0<2.62$, according to this prescription, one cannot
construct an interval containing probability $>90\%$ that does not include
$n=0$ and we adopt 90\% one-sided intervals.

\bibitem{bib:belesev} A. I. Belesev, Phys. Lett. {\bf B350}, 263 (1995).

\bibitem{fn:3}To compute the intervals [$x_1,x_2$] of Ref.  \cite{bib:fc}
at significance $\alpha$ for small $\mu$, we use $\mu=x_2^2/(2(x_2-x_1))$
and the approximation $erf(x_2/\sqrt 2) \simeq 2\alpha-erf(-x_1/\sqrt 2)$. 

\end{thebibliography}
\end{document}